\documentclass[12pt]{article}

\usepackage{amssymb}
\usepackage{amsmath}
\usepackage{latexsym}
\usepackage{yfonts}

\catcode`@=11
\renewcommand{\section}{\@startsection{section}{1}{0pt}{\medskipamount}
{\medskipamount}{\large\bf}} \numberwithin{equation}{section}
\catcode`@=12

\def\beq{\begin{eqnarray}}    
\def\eeq{\end{eqnarray}}      

\def\={\ =\ }

\begin{document}
\begin{center}

{\Large\bf On quantum-mechanical equations of motion
in representation dependent of external sources
}

\vspace{18mm}

{\large Igor A. Batalin$^{(a,b)}\footnote{E-mail:
batalin@lpi.ru}$\;, Peter M. Lavrov$^{(b, c)}\footnote{E-mail:
lavrov@tspu.edu.ru}$\; }

\vspace{8mm}

\noindent ${{}^{(a)}}$
{\em P.N. Lebedev Physics Institute,\\
Leninsky Prospect \ 53, 119991 Moscow, Russia}

\noindent  ${{}^{(b)}} ${\em
Tomsk State Pedagogical University,\\
Kievskaya St.\ 60, 634061 Tomsk, Russia}

\noindent  ${{}^{(c)}} ${\em
National Research Tomsk State  University,\\
Lenin Av.\ 36, 634050 Tomsk, Russia}

\vspace{20mm}

\begin{abstract}
\noindent In the present paper, we consider in detail  the aspects
of the Heisenberg's equations of motion, related to their
transformation to the representation dependent of external sources.
We provide with a closed solution as to the variation-derivative
motion equations in the general case of a normal form (symbol)
chosen. We show that the action in the path integral does depend
actually on a particular choice of a normal symbol. We have
determined both the aspects of the latter dependence: the specific
boundary conditions for virtual trajectories, and the specific
boundary terms in the action.

\end{abstract}

\end{center}

\vfill

\\

\newpage

\section{Introduction}

In the present paper, we consider in detail the general scenario of
quantum-mechanical dynamical description as originated by
Heisenberg, Schroedinger, Dirac, Schwinger, Dyson, DeWitt
\cite{Schroed,Heis,Feynman1,Feynman2,Dirac,Sch,Schwinger,Dy2,Dy3,DeWitt,DeW1,DeW},
and then developed essentially by Fradkin,  Faddeev
\cite{Fradkin1,Fradkin,Faddeev}, and  members of their scientific
schools
\cite{FP,FTyu,S,FaddeevPop,BF1975,T,FVh1,FVh2,BVh,FF,FaddSlav,
VLT,BFr1,BF2,BB2013,BB2014}.

We begin with the standard Heisenberg's equations of motion as
formulated for an autonomous dynamical system, and then we define
the transformation operator generating a new representation where
dynamical equations look as formulated in the presence of external
sources. We show that the generating operator introduced appears to
be, at the same time, a generating functional (of the external
sources) for all chronologic - ordered products of operators in the
new representation. On the ground of that observation, we
reformulate the Heisenberg's equation of motion in terms of
chronologic - ordered operator products via the respective time
delocalization as for the normal symbol of the Hamiltonian operator
\cite{vonNeum,Weyl,Ber,Berezin,BerShub}.
In this way, we derive closed variation - derivative equations as
for the generating operator of chronologic - ordered operator
products.

We show that the variation-derivative equations derived are
integrable in the sense of the compatibility criterion, due to the
canonical commutation relations as for the equal - time operators.
In this way, we find ourselves allowed to integrate the variation
derivative -equations along the straight line,  as to obtain their
closed solution when using the Weyl's normal form (symbol).  Then,
by making use of the Baker - Campbell - Hausdorff formula, one can
extend the solution obtained as to the cases of the $QP$ or the $PQ$
normal symbols.

By applying the functional Fourier transformation as to the
generating operator in its dependence on the external sources, one
derives the path - integral form of the solution to the
variation-derivative equations of motion. It appears that the action
in the path integral depends actually on the particular normal form
(symbol) chosen \cite{Berezin,BerShub}. We consider the two types of
the path integral. The first kind, called "general", does not assume
the phase variables to be split explicitly into co-ordinates and
momenta. Vice versa,  the second kind of the path integral, called
"split", does. For both the kinds, we provide with convenient ways
as to how to parameterize the dependence of the action on the
particular choice of normal symbol. The latter dependence has the
two aspects: the specific boundary conditions for virtual
trajectories, and the specific form of the boundary terms in the
action. We have determined both these aspects. It is also a
remarkable feature that a restricted virtual trajectory is expressed
in terms of an unrestricted velocity, being the latter an actual
path integration variable. In the latter sense, our path integrals
represent a quantization of the field of velocities.

Finally, we extend the above results as to their form symmetric in
the chronologic and the anti-chronologic ordering. With this
purpose, we introduce the second kind of the external source, and
then modify appropriately the equation  defining the generating
operator.
We provide
with a closed variation-derivative solution symmetric in the
chronologic and the anti-chronologic operator ordering.
\\

\section{Heisenberg's equations of motion and generating operator
for chronologic products}

Let $Z^{A}( t )$, $\varepsilon(Z^{A})=:\varepsilon_A$, be canonical phase-variable operators, and $H( Z )$
be an original Hamiltonian, so that the Heisenberg's equations of
motion read
\beq
\label{H1.1}
i \hbar \frac{ \partial  Z^{ A } }{
\partial t }  =  [  Z^{ A },  H ], \quad
[  Z^{ A },  Z^{ B }  ]  =  i \hbar \omega^{ AB }  =  {\rm const}.    
\eeq
Let $J_{A}( t )$, $\varepsilon(J_{A})=:\varepsilon_A$, be an external source,
and let us define a
transformation operator $U( t )$ as to satisfy the equation
\beq
\label{H1.2}
i \hbar \frac{ \partial  U }{ \partial t }  =  -  J_{ A }  Z^{ A } U,
\quad U( - \infty )  =  1.       
\eeq
Now, we define canonical phase-variable operators in the
source-dependent representation,
\beq
\label{H1.3}
Z'^{ A }  =:  U^{ -1}  Z^{ A }  U,  
\eeq
as to satisfy the modified equation of motion
\beq
\label{H1.4}
i \hbar \frac{ \partial Z'^{ A } }{ \partial t }  =  [  Z'^{ A },
H' -  J_{ B }  Z'^{ B }  ].    
\eeq
It follows from (\ref{H1.2}) that
\beq
\label{H1.5} i \hbar
U^{-1}( t ) \frac{ \partial }{ \partial t }  \frac{ \delta }{ \delta
J_{ A }( t' ) } U( t )  =- \delta ( t - t' )  Z'^{A}( t )  - J_{ B
}( t )  Z'^{ B }( t )  U^{-1}( t )  \frac{ \delta }{ \delta
J_{ A }(  t' ) }  U( t ). 
\eeq
On the other hand, the left-hand side in (\ref{H1.5}) rewrites as
\beq
\label{H1.6}
i \hbar  \frac{ \partial }{ \partial t }  U^{-1}( t ) \frac{ \delta
}{ \delta J_{ A }( t' ) }  U( t )  -
\left(  i \hbar \frac{ \partial }{ \partial t } U^{-1}( t )  \right)  \frac{
\delta }{ \delta J_{ A }( t' ) }  U( t ).  
\eeq
Here in (\ref{H1.6}), in the second term,  the first parentheses rewrite as
\beq
\label{H1.7}
i \hbar \frac{ \partial }{ \partial t } U^{-1}( t )  =
 -  U^{-1}( t )  \left(  i \hbar  \frac{ \partial }{\partial t }  U( t
)  \right)  U^{-1}( t )  =
J_{ B }( t )  Z'^{ B }( t )  U^{-1}( t ).    
\eeq
In this way, the second term in (\ref{H1.6}) cancels exactly the second term
in the right-hand side in (\ref{H1.5}), and the latter (\ref{H1.5}) takes the form
\beq
\label{H1.8}
i \hbar  \frac{ \partial }{ \partial t }  U^{-1}( t )  \frac{ \delta
}{ \delta J_{ A }( t' ) }  U( t )  =
-  \delta( t - t' )  Z'^{A}( t' ),  
\eeq
whose causal solution reads
\beq
\label{H1.9}
\frac{ \hbar }{ i }  U^{ -1 }( t )  \frac{ \delta }{ \delta J_{ A }(
t' ) }  U( t )  =
 \theta( t - t' )  Z'^{ A }( t' ),    
\eeq
so that
\beq
\label{H1.10}
\frac{ \hbar }{ i }  \frac{ \delta }{ \delta J_{ B }( t' ) } Z'^{ A
}( t )  =
[  Z'^{ A }( t ),  \theta( t - t' )  Z'^{ B }( t' )  ]  (-1)^{
\varepsilon_{ A } \varepsilon_{ B } }.    
\eeq
In the limit $t\rightarrow  \infty$,  it follows from (\ref{H1.9}), (\ref{H1.10})
that
\beq
\label{H1.11}
\left( \frac{ \hbar }{ i } \right)^{ 2 } \frac{ \delta^{ 2 } }{ \delta J_{ B }(
t^{''} ) \delta J_{ A } ( t' ) }\;  U( \infty )  =
U( \infty ) \; T (  Z'^{ B }( t^{''} )  Z'^{ A }( t' )  ),   
\eeq
where the binary chronologic-ordered product is defined by
\beq
\label{H1.12}
T (  Z'^{ B }( t^{''})  Z'^{ A }( t' )  )  =:  \left(  \theta( t^{''} - t' )
Z'^{ B }( t^{''})  Z'^{ A }( t' )  +
\theta( t' - t^{''} )  Z'^{ A }( t' ) Z'^{ B }(t^{''})  (-1)^{ \varepsilon_{
B } \varepsilon_{ A } }  \right),  
\eeq
see also (\ref{H1.18}) below for the general case.

Let $F[ Z ]$ be a functional of classical functions $Z^{ A }( t )$,
and \beq \label{H1.13}
F[0]=0,\quad F_{ A }( t | Z )  =:  F[ Z ]
\frac{ \overleftarrow{\delta } }{ \delta
Z^{ A }( t ) }.       
\eeq
By applying the variation-derivative operator,
\beq
\label{H1.14}
F\left[ \frac{ \hbar }{ i }  \frac{ \delta }{ \delta J( \cdot ) }  \right],  
\eeq
as to the equation (\ref{H1.2}), we get
\beq
\label{H1.15}
\left(  i  \hbar \frac{ \partial }{ \partial t }  +
J_{ A }  Z^{ A }  \right)  F\left[ \frac{  \hbar }{ i }  \frac{ \delta }{
\delta J( \cdot ) }  \right]  U   =
 -  \frac{ \hbar }{ i }  F_{ A }\left( t | \frac{ \hbar }{ i }  \frac{
\delta }{ \delta J( \cdot ) }  \right)  Z ^{ A }  U.  
\eeq
It follows from (\ref{H1.15}) that
\beq
\nonumber
F\left[ \frac{ \hbar }{ i }  \frac{ \delta }{ \delta J( \cdot ) } \right ]  U( t
)  &=&   \int_{ - \infty }^{ t }  dt'  U( t )  U^{ -1 }( t' )
F_{ A }\left(  t' | \frac{ \hbar }{ i }  \frac{ \delta }{ \delta J( \cdot
) }  \right)  Z^{ A }( t' )  U( t' )  =\\
\label{H1.16}
&=&
 U( t )  \int_{ - \infty }^{ t }  dt'  F_{ A }\left(  t' | \theta( t' -
( \cdot )  )  Z'( \cdot )  +
 \frac{ \hbar }{ i } \frac{ \delta }{ \delta J ( \cdot ) }  \right)
Z'^{A}( t' ).    
\eeq
Here in (\ref{H1.16}), in the second equation, we have used
the above eq.(\ref{H1.9}). If  one chooses a functional $F[ Z ]$ in the monomial form,
\beq
\label{H1.17}
F[ Z ]  =  Z^{ A_{ 1} }( t_{ 1 } ) \; \cdots \; Z^{ A_{ n } }( t_{ n} ), 
\eeq
then, in the limit  $t \rightarrow \infty$,  the right-hand side in (\ref{H1.16})
does
generate the corresponding $T$-product as to the primed  operators (\ref{H1.3})
in the external source-dependent representation,
\beq
\label{H1.18}
\frac{ \hbar }{ i } \frac{ \delta }{ \delta J_{ A_{ 1 } }( t_{ 1 } )
} \; \cdots\;
\frac{ \hbar }{ i } \frac{ \delta }{ \delta J_{ A_{ n} } ( t_{ n } )
} \; U( \infty )  =
U( \infty )\; T (  Z'^{ A_{ 1} }( t_{ 1 } )\; \cdots\; Z'^{ A_{ n } }(
t_{ n } )  ).    
\eeq

\section{Variation - derivative equations of motion
for generating operator and their closed solution}

Now, let us represent the primed Hamiltonian $H'$ in the form of its chronologic-ordered
delocalized symbol $H_{ R }$,
\beq
\label{H1.19}
H' =  H( Z'( t ) )  =  T \left(  H_{ R }( t | Z'( \cdot ) )  \right),    
\eeq
and the same for the commutator
\beq
\label{H1.20}
[  Z'^{ A }( t ),  H( Z'( t ) )  ]  =
i \hbar  \omega^{ AB } T \left(  \frac{ \delta }{ \delta Z'^{ B }( t ) }
\int_{ - \infty }^{ \infty }  dt'  H_{ R }( t' | Z'( \cdot ) )  \right).   
\eeq
Then, due to the eq.(\ref{H1.18}), it follows from the primed Heisenberg's
equation (\ref{H1.4}) that the
variation-derivative equation of motion holds as for the operator $U( \infty )$,
\beq
\label{H1.21}
\left[  \dot{ Z }^{ A }( t )  -  \omega^{ AB }  \left(  \frac{ \delta }{ \delta Z^{ B }( t ) }
\int_{  -  \infty }^{  \infty }
dt'  H_{ R }( t' | Z( \cdot ) )  \right)  -
J_{ B }( t )\; \omega^{ BA }  \right]\!\! \left( Z  =
 \frac{ \hbar }{ i }  \frac{ \delta }{ \delta J }\right) \; U( \infty )  =  0.      
\eeq
A variation-derivative solution to that equation has the form \cite{Wick,Berezin}
\beq
\nonumber
U( \infty )  &=&
\exp\left\{  -  \frac{ i }{ \hbar }  \int_{-\infty }^{ \infty }  dt
H_{R}\left(t|\frac{\hbar }{ i }\frac{\delta }{\delta J (\cdot )}\right)\right\}\times\\
\nonumber
&&\times\exp\left\{  -  \frac{ i }{ \hbar }  \frac{ 1 }{ 2 }
\int_{ - \infty }^{ \infty } \!\! dt  \int_{ - \infty }^{ \infty } \!\! dt'
J_{ A }( t )  \mathcal{ D }^{ AB }( t - t' )  J_{ B }(  t' )
( -1 )^{ \varepsilon_{ B } }\right\}\times\\
\label{H1.22}
&&
\times N \left(  \exp\left\{  \frac{ i }{ \hbar }
\int_{ - \infty }^{ \infty }  dt  J_{ A }( t )  Z'^A_{ in }  \right\}  \right),    
\eeq
\beq
\label{H1.23} \mathcal{ D }^{ AB }( t - t' )  = :  \frac{
1 }{ i \hbar } \left( Z'^{ A }_{in}  Z'^{ B }_{in}\;\!\theta(t-t') +
Z'^{ B }_{in}  Z'^{ A
}_{in}\;\!\theta(t'-t)(-1)^{\varepsilon_A\varepsilon_B}-
N (  Z'^{ A }_{in}  Z'^{ B }_{in}  )  \right),    
\eeq
\beq
\label{H1.24}
\frac{ \partial }{ \partial t } \mathcal{ D }^{ AB }( t - t' )
=  \omega^{AB}  \delta( t - t' ),    
\eeq
\beq
\label{H1.25}
Z'^A_{ in }  =:  \frac{ \hbar }{ i } \; U^{-1}( \infty )\;
\frac{ \delta U( \infty ) }{ \delta J_{ A }( - \infty ) }.  
\eeq

In order to explain as to  how to derive the solution (\ref{H1.22}),
let us seek for a solution to (\ref{H1.21}) in the natural form
\beq
\label{H1.26a} U( \infty )  =  \exp\left\{  -  \frac{ i }{ \hbar }
\int_{ - \infty }^{ \infty } dt\;\! H_{ R }\left(  t | \frac{ \hbar
}{ i }  \frac{ \delta }{ \delta J( \cdot ) }
\right)  \right\} U_{ ext }( \infty ),    
\eeq
so that the operator $U_{ ext }( \infty )$ does satisfy the equation
\beq
\label{H1.27a}
\left[  \frac{ \partial }{ \partial t }  \frac{ \hbar }{ i }
\frac{ \delta }{ \delta J_{ A }( t ) }  +
\omega^{ AB } J_{ B }( t )  (-1)^{ \varepsilon_{ B } }  \right]
U_{ ext }(\infty)  =  0,
\eeq
or in its integral form,
\beq
\label{H1.28a}
\frac{ \hbar }{ i }  \frac{ \delta U_{ ext }( \infty ) }{ \delta J_{A }( t ) }  =
U_{ ext }( \infty )
\left(  Z_{ in }'^{ A }  -
\int_{ - \infty }^{ t }  dt'  \omega^{ AB }  J_{ B }( t' )
(-1)^{ \varepsilon_{ B } }  \right) ,    
\eeq
with
\beq
\label{H1.29a}
Z'^{ A }_{ in }  =:  Z'^{ A }( - \infty ),    
\eeq
being  $in$ - operators independent of the  external sources $J_{A}( t )$.
The equations (\ref{H1.28a})
are integrable due to the canonical commutation relations as for the operators
(\ref{H1.29a}); we have from (\ref{H1.28a})
\beq
\nonumber
&&X^{ BA }( t', t )  =:
\left(  \frac{ \hbar }{ i }  \right)^{ 2 }\frac{  \delta^{ 2 }  U_{ ext }( \infty )  }
{  \delta J_{B }( t')  \delta J_{ A }( t )  }  =
U_{ ext }( \infty )
\left[\left(  Z'^{ B }_{ in }  -  \int_{ - \infty }^{ t' }
ds' \omega^{ BD }  J_{ D }( s' )  (-1)^{ \varepsilon_{ D } }  \right)\right.\times\\
\label{H1.30a}
&&\qquad\qquad\qquad\qquad\left.\times\left(Z'^{ A }_{ in }  -\int_{ - \infty }^{ t }
ds \;\! \omega^{ AC }  J_{ C }(  s  ) (-1)^{ \varepsilon _{ C } }  \right)  +
\frac{ \hbar }{ i } \omega^{ BA } \theta( t - t' )  \right] .    
\eeq
By anti - symmetrizing the operators ( \ref{H1.30a}), we get
\beq
\nonumber
&&X^{ BA }( t', t )  -
X^{ AB }( t, t' ) (-1)^{ \varepsilon_{ B }  \varepsilon_{ A } }  =\\
\label{H1.31a}
&&=U_{ ext }( \infty )
\left(  [  Z'^{ B }_{ in },  Z'^{ A }_{ in }  ]  +
\frac{ \hbar }{ i }  \left(  \omega^{ BA } \theta( t - t' )  -
\omega^{ AB }  \theta( t' -  t )
(-1)^{ \varepsilon_{ B } \varepsilon_{ A } }  \right)  \right) .       
\eeq
As  the $in$ - operators do satisfy the canonical commutation relations,
the  commutator in the
right-hand side in (\ref{H1.31a}) equals to $i \hbar\;\! \omega^{BA }$,
while the intrinsic parentheses equal to
\beq
\label{H1.32a}
\omega^{ BA }(  \theta( t - t' )  +  \theta( t' - t )  )  =  \omega^{ BA }.  
\eeq
Thus, we have shown that the anti - symmetric part of the operators (\ref{H1.30a})
is zero,
\beq
\label{H1.33a}
X^{ BA }( t', t )  -  X^{ AB }( t, t')
( -1)^{ \varepsilon_{ B } \varepsilon_{ A} }  =  0,   
\eeq
which means that the integrability holds \cite{Bat} as to the equation (\ref{H1.28a}).
By multiplying
the latter (\ref{H1.28a}) by $\delta J_{ A }( t )$ from the left,
and integrating then over $dt$, we
get the complete variation as for the operator $U_{ ext }( \infty )$,
\beq
\label{H1.34a}
\frac{ \hbar }{ i } \delta U_{ ext }( \infty )  =
U_{ ext }( \infty )  \int_{ - \infty }^{ \infty}  dt\;\!
\delta J_{ A }( t )  \left(  Z'^{ A }_{ in }  -
\int _{ - \infty }^{ t }  dt'  \omega^{ AB }  J_{ B }( t' )
(-1)^{ \varepsilon_{ B } }  \right) .    
\eeq
By making a formal rescaling
\beq
\label{H1.35a}
J_{ A }( t ) \; \rightarrow\;  \lambda J_{ A }( t ),    
\eeq
with $\lambda$ being a Boson parameter, due to the integrability (\ref{H1.33a})
shown above,
we are allowed to choose the variation of the rescaled source
 along the straight line,
\beq
\label{H1.36a}
\delta \big(\lambda J_{ A }( t )\big)  =  d \lambda  J_{ A }( t ).   
\eeq
Then we get the differential equation in $\lambda$,
\beq
\nonumber
\frac{ \hbar }{ i }  \frac{ \partial U_{ ext }( \infty ) }{ \partial \lambda }  &=&
U_{ ext }( \infty )\left[  \int_{ - \infty }^{ \infty }
dt \;\!J_{ A }( t )Z'^{ A }_{ in }\;-\right.\\
\label{H1.37a}
&&\left.-\lambda  \int_{ - \infty }^{ \infty }  dt
\int_{ - \infty }^{ \infty }  dt'
J_{ A }( t ) \;\! \omega^{ AB }  \frac{ 1 }{ 2 }\;\!
{\rm sign}( t - t' )  J_{ B }( t )  (-1)^{ \varepsilon_{ B } }  \right]\!. 
\eeq By integrating that equation together with the condition
$U_{ ext }( \infty ) |_{ \lambda  =  0 }  =  1$, and taking then $\lambda  = 1$,
we arrive finally at the solution for the $U_{ ext }( \infty )$ as
given by the product of the second and the third exponential in
(\ref{H1.22}),  with $N$ being the Weyl normal form;
other types of normal form for $N$ do follow via the Baker - Campbell
- Hausdorff invariant formula,
\beq
\nonumber
\exp\Big\{\frac{ i }{ \hbar }
\int_{- \infty }^{ \infty }\! dt J_{ A }( t )  Z^{'A}_{ in }  \Big\}\!\! & =&\!\!
\exp\Big\{ \!\! -  \frac{ i }{ 2 \hbar}  \int_{ - \infty }^{ \infty } \!\!\! dt
\int_{ - \infty }^{ \infty } \!\!\! dt'
J_{ A }( t )  \frac{ 1 }{ i \hbar }  ( N_{ W } - N )
( Z^{'A}_{ in }  Z^{'B}_{ in } )  J_{ B }( t' )  (-1)^{ \varepsilon_{ B }}\Big\}\!\!\times\\
\label{H1.38a}
&&\times N \Big( \exp\Big\{  \frac{ i }{ \hbar }  \int_{ - \infty }^{ \infty }  dt
J_{ A }( t )  Z^{'A}_{ in } \Big\}\Big). \!\!\!\!  
\eeq

\section{Path-integral solution for generating operator}

The path integral solution can easily be derived from (\ref{H1.22})
by making use of a functional
Fourier transformation as for the second exponential.
Let us split temporarily the complete set of canonical phase-variable  operators
$Z^{ A }$
into the sectors of co-ordinates $Q^{ i }$ and momenta $P_{ i }$,
\beq
\label{H1.26}
Z^{ A }  =:  ( Q^{ i } ; P_{ i } ).  
\eeq
Denote via $N_{ (a, b) }$,  $a + b = 1$,   the $( a, b )$ - type normal form:
\beq
\label{H1.27}
&&a  =  0, \quad   b  =  1 :  \quad    QP, \\  
\label{H1.28}
&&a  =  1,  \quad  b  =  0 :   \quad   PQ, \\    
\label{H1.29}
&&a  =  \frac{ 1 }{ 2 }, \quad   b  =  \frac{ 1 }{ 2 } :  \quad   Weyl.   
\eeq
Then, we have
\beq
\label{H1.30}
H_{ R }( t | Z( \cdot ) )  =  H_{(a,b)}( Z_{ R }( t ) ),   
\eeq
where the delocalized set of phase variables is given explicitly by
\beq
\label{H1.31}
Z_{ R }^{ A }( t )  =:  ( Q^{ i }( t + {\rm sign}( b - a )\;\! 0 ) ;
P_{ i }( t )  ). 
\eeq
As for the cases (\ref{H1.27}) - (\ref{H1.29}),
the $\mathcal{ D }$ function (\ref{H1.23})  rewrites in the form \cite{BFrad}
\beq
\label{H1.32}
\mathcal{ D }^{AB}( t - t' )  =\frac{1}{2}{\rm sign}(t-t')\;\!\omega^{AB}+
\frac{1}{i\hbar}\left(N_{(\frac{1}{2},\frac{1}{2})}-N_{(a,b)}\right)
(Z'^A_{in}Z'^B_{in}).
\eeq
Thus, the $( a, b )$ - type {\it in'} - symbol $U_{ (a, b) }( \infty )$
of the operator $U( \infty )$ is
given by the path integral
\beq
\label{H1.33}
U_{ (a, b) }( \infty )=
\int [ DV ]  \exp\left\{\frac{ i }{ \hbar }W\right\}, 
\eeq
where the measure $[ DV ]$ contains the required normalization factor as to maintain
that (\ref{H1.33}) equals to one at $J = 0$,  while the action $W$ has the form
\beq
\nonumber
&&\!\!\!\!\!W =: \! \int_{ - \infty }^{ \infty } \!\! dt
\left[  \frac{ 1 }{ 2 }  Z^{ A}( t ) \;\! \omega_{ AB}  \dot{ Z }^{ B }( t )  -
H_{ (a, b) }( Z_{ R }( t ) )  +  J_{ A }( t )  Z^{ A }( t )  \right] +\\
\label{H1.34}
&&\!\!\!\!\!+\frac{1}{2}\;\! Z^{ A }(\infty ) \; \omega_{ AB }\;\!  Z^{ B }(- \infty )
+\frac{1}{2}\big(P_i(\infty)+P_i(-\infty)-2P'_{i\;in}\big)\big(Q^i(\infty)-Q^i(-\infty)\big),
\eeq
with  $Z^{A}( t )$ being a restricted virtual trajectory,
\beq
\nonumber
&&\frac{1}{2} \left(Z^{ A }( \infty )  +   Z^{ A }( - \infty )\right)
-\\
\label{H1.35}
&&-\frac{1}{i\hbar}
\left(N_{(\frac{1}{2},\frac{1}{2})}-N_{(a,b)}\right)(Z'^A_{in}Z'^B_{in})\;\omega_{BC}
\left(Z^{ C }( \infty )  -   Z^{ C }( - \infty )\right)
=  Z'^A_{ in },   
\eeq as expressed via an unrestricted velocity $V^{ A }( t )$, \beq
\label{H1.36} Z^{A}( t )  =:  Z'^A_{ in }  +  \int_{ - \infty }^{
\infty }  dt'\;
\mathcal{ D }^{AB}( t - t' )\;\!\omega_{BC}\;\!  V^{ C}( t' ).  
\eeq In this way, the above path integral (\ref{H1.33}) appears to
be, in fact, an actual quantization of the field of velocities. It
follows from the general boundary conditions (\ref{H1.35}) that
there holds their split version in the form
\beq
\label{H1.36a}
&&b \;\!Q^{ i }( \infty )  +  a\;\! Q^{ i }( - \infty  )  =  Q'^{ i }_{ in },\\ 
\label{H1.36b}
&&a P_{ i }( \infty  )  +  b P_{ i }( - \infty  )  =  P'_{ i \; in } . 
\eeq

In order to derive in a natural way the path integral
(\ref{H1.33}) from the variation-derivative solution (\ref{H1.22}),
one may notice the path-integral Fourier representation for the
$(a, b)$-type $in'$-symbol of the operator $U_{ ext }( \infty )$,
\beq \label{H1.49} U_{ ext (a, b) }( \infty )  =
\int [ DV ]  \exp\left\{  \frac{ i }{ \hbar }  W_{ ext }  \right\},   
\eeq
\beq
\nonumber
&&\!\!\!\!\!W_{ ext }  =:  \int_{ - \infty }^{ \infty }  dt  \left[ \frac{1}{2} Z^{ A }( t )\;\!
\omega_{ AB }  \dot{ Z }^{ B }( t )  +
 J_{ A }( t ) Z^{ A }( t )  \right]  +\\
 \label{H1.50}
 &&\!\!\!\!\!+\frac{1}{2}\;\! Z^{ A }(\infty ) \; \omega_{ AB }\;\!  Z^{ B }(- \infty )
+\frac{1}{2}\big(P_i(\infty)+P_i(-\infty)-2P'_{i\;in}\big)\big(Q^i(\infty)-Q^i(-\infty)\big),
\eeq
where the formulae (\ref{H1.32}), (\ref{H1.35}), (\ref{H1.36})
must be taken into account. As the action (\ref{H1.50}) is quadratic in $Z^{A}( t )$,
the path integral (\ref{H1.49}) is a Gaussian one.
Therefore, its value is given by taking the Boltzmann factor,
$\exp\left\{ \frac{ i }{ \hbar } W_{ext} \right\}$,
at the extremum of the action (\ref{H1.50}), which is given by
\beq
\label{H1.50a}
Z^{ A }(t)  =  Z'^{ A }_{ in }  -  \int_{ -\infty }^{ \infty }  dt'\;
\mathcal{D}^{AB}( t - t' )\;\!  J_{B }( t' )  (-1)^{ \varepsilon_{ B } } .    
\eeq
In this way, one reproduces immediately the product of the second and the third
exponential in (\ref{H1.22}).

 Now, let us split the set of external sources in accordance with the splitting
(\ref{H1.26}) as for the phase variables
\beq
\label{H1.51}
J_{ A }  =: (  I_{ i } ;  K^{ i }  ).  
\eeq
Then, the variation-derivative solution (\ref{H1.22}) takes the split
form,
\beq
\nonumber
&&U( \infty )  =\;\! \exp\left\{  -  \frac{ i
}{ \hbar }  \int_{ - \infty }^{ \infty }  dt\;\! H_{ R }\left( t |
\frac{ \hbar }{ i }  \frac{ \delta }{\delta I( \cdot )} ;
\frac{ \hbar }{ i }  \frac{ \delta }{ \delta K( \cdot ) } \right)  \right\}\times \\
\nonumber &&\qquad\quad\;\!\times \exp\left\{ - \frac{ i }{ \hbar }
\int_{ - \infty }^{ \infty } dt \int_{ - \infty }^{ \infty }  dt'
I_{ i }( t )\;\! \mathcal{ D }^{ i }_{ \;j }(t-t')\;\! K^{ j }( t' )
(-1)^{ \varepsilon_{ j } }\right\}\times\\
\label{H1.52} &&\qquad\quad\;\!\times N\left(\exp\left\{  \frac{ i }{ \hbar}
\int_{ - \infty }^{ \infty }  dt\;\!
( I_{ i }( t )  Q'^{ i }_{ in }  +  K^{ i }( t )  P'_{ i\;\! in }  )  \right\}\right) ,     
\eeq
where
\beq
\nonumber
\mathcal{ D }^{ i }_{ \;j }( t - t' )  &=:&  \frac{ 1 }{ i \hbar }
\left(
Q'^{i}_{in}  P'_{j\; in}  \theta( t - t' )  +
P'_{j\; in}  Q'^{i}_{in}  \theta( t' - t )  (-1)^{ \varepsilon_{i} \varepsilon_{j} }  -
N (  Q'^{ i }_{in}  P'_{ j\; in })\right) =\\
\label{H1.53}
& =&
\delta^{ i }_{ \;j }
\big(  a\;\! \theta( t - t' )  -  b\;\! \theta( t' -  t )  \big). 
\eeq In a similar way we get for the transposed $\cal{D}$-function,
\beq
\label{H1.53a}
\mathcal{ D }^{ \;\;i }_{ j }( t - t' )=
-(-1)^{\varepsilon_i\varepsilon_j}\delta^{\;\; i }_{ j } \big( b\;\!
\theta( t - t' )  -  a\;\! \theta( t' -  t )  \big). \eeq
The non-split function (\ref{H1.23})/(\ref{H1.32})
rewrites in terms of the blocks (\ref{H1.53}),
(\ref{H1.53a})
\beq
\mathcal{ D }^{ AB }( t - t' )  =
\begin{pmatrix}
0&\mathcal{ D }^{i }_{ \;j }( t - t' )\\\mathcal{ D }^{ \;\;i }_{ j
}( t - t' )&0
\end{pmatrix}
.
\eeq

Introduce the following path - integral Fourier representation
\beq
\nonumber
&&\exp\left\{  -  \frac{ i }{  \hbar }  \int_{ - \infty }^{ \infty }  dt
\int_{ - \infty }^{ \infty }  dt'
I_{ i }( t ) \;\! \mathcal{ D }^{ i }_{ \;j }( t - t' )
K^{ j }( t' )  (-1)^{ \varepsilon_{ j } }  \right\}  =\\
\nonumber &&= \int  [ DP ]  [DV ]  \exp\left\{  \frac{ i }{ \hbar }
\int_{ - \infty }^{ \infty }  dt \Big[ P_{ i }( t )  \big(  V^{ i }(
t )  +  K^{ i }( t )
(-1)^{ \varepsilon_{ i } }  \big)  +\right.\\
\label{H1.54} &&\left.\qquad\qquad\qquad\qquad\;\;\;\; +I_{ i }( t )
\!\!\int_{ - \infty }^{ \infty }  dt'
\mathcal{ D }^{ i }_{ \;j }( t - t' )  V^{ j }( t' )  \Big]  \right\}.   
\eeq
By making use of that representation,  we arrive finally at the following split
path-integral formula as for the $(a, b)$-type $in'$-symbol of the operator
$U( \infty )$,
\beq
\nonumber
&&U_{ (a, b) }( \infty )=  \int  [ DP ]  [ DV ]
\exp\left\{  \frac{ i }{ \hbar }   \int _{ - \infty }^{ \infty }\!\!\!  dt\!
\Big[  P_{ i }( t ) \;\! \dot{ Q }^{ i }( t )  -
H_{(a,b)}(  Q_{ R }( t ), P_{ R }( t ))\;  +
\right.\\
\label{H1.55}
&&\qquad\qquad\qquad + \;I_{ i }( t )\;\!Q^{ i }( t )  +
 K^{ i }( t )P_{ i }( t )\Big]\left. -
\frac{ i }{ \hbar }  P'_{i\; in  }
\big(  Q^{ i }( \infty )  -  Q^{ i }( - \infty )\big)\right\}, 
\eeq where \beq \label{H1.56b}
b\;\!Q^i(\infty)+a\;\!Q^i(-\infty)=Q'^i_{in},
\eeq
\beq
\label{H1.56} Q^{ i }( t )  =:  Q'^{ i }_{ in }  +
\int_{  -  \infty }^{  \infty } dt'\;\! \mathcal{ D }^{ i }_{ \;j }( t - t') V^{ j }(t').
\eeq Here in (\ref{H1.55}), the $Q_{ R }$ and $P_{ R }$ are just
given explicitly in components in the right-hand side in
(\ref{H1.31}). In the split path integral (\ref{H1.55}), the momenta
$P_{ i }( t )$ are non-restricted integration variables, so that
there are no boundary conditions to them. However, consider the
classical equations of motions for velocities $V^{ i }( t )$, as
they follow from the action in (\ref{H1.55}),
\beq
\label{H1.56a}
P_{ i }( t )  =  P'_{ i \; in }  + \Big(  a \int_{ t }^{ \infty }  -
b \int_{ - \infty }^{ t }  \Big)\;  dt'
\;\frac{ \partial H_{ ( a, b ) } }{ \partial Q^{ i }( t' ) } .     
\eeq
It follows from (\ref{H1.56a})
\beq
\label{H1.56b}
\dot{P}_{ i }( t )  =  - \frac{ \partial H_{ ( a, b) } }{ \partial Q^{ i }( t ) },
\eeq
\beq
\label{H1.56c}
&&a P_{ i }( \infty )  +  b P_{ i }( - \infty )  =  P'_{ i \; in } .   
\eeq
Thus, in the split path integral (\ref{H1.55}), we have reproduced the boundary conditions
(\ref{H1.36b}) for momenta $P_{ i }( t )$ at the classical level.

As to compare our proposed path integrals for the ones of Berezin and Shubin, the
situation is the following. In the book of these authors \cite{BerShub},
the general (non-split)
path integral, similar to our (\ref{H1.33}), is given as applied only
to the case of the Weyl
symbol. In the cases of the $PQ$ and the $QP$ symbol, these authors have given only
split path integrals, similar to our (\ref{H1.55}).
It should also be mentioned that all our
path integrals have just the respective unrestricted velocities as their actual
integration variables, rather than restricted trajectories.

\section{Formulation symmetric in chronologic and anti-chronologic\\
products}

So far, we did proceed from the basic idea of the chronologic ordering
of operators. However, it is a remarkable feature that there also exists
an approach \cite{Keldysh,Marinov} based on symmetric use of both the chronologic and the
anti - chronologic product. In that approach, one proceeds with extended
version of the equation (\ref{H1.2}),
\beq
\label{H1.58}
i \hbar  \frac{ \partial U }{ \partial t }  =   -  J_{ A }  Z^{ A }  U  +
U \tilde{ J }_{ A}  Z^{ A}, \quad  U( - \infty )  =  1  
\eeq
where $Z^{ A }$  are the Heisenberg's operators as defined by (\ref{H1.1}),
and  $\tilde{ J }_{ A }( t )$ is
a new external source.
The equation (\ref{H1.58}) is satisfied naturally with the factorized ansatz
\beq
\label{H1.58a}
U  =  U_{ J }  (U_{ \tilde{ J } })^{-1},      
\eeq
where the operator $U_{ J }$ is defined by (2.2),
\beq
\label{H1.58b}
i \hbar \frac{ \partial  U_{ J } }{ \partial t }  =
-  J_{ A }  Z^{ A }  U_{ J }, \quad U_{ J }( - \infty )  =  1, 
\eeq
and the operator $U_{ \tilde{ J } }$ is defined by  (\ref{H1.58b})
with the replacement $J_{ A } \rightarrow \tilde{ J }_{ A }$.

Then, by making use of the same method as the one
applied as to the  eq.(\ref{H1.2}), one can derive easily
a variation - derivative solution
as to the equation (\ref{H1.58}),
\beq
\nonumber
&&\!\!\!U( \infty )  =
\exp\left\{  -  \frac{ i }{ \hbar } \int_{ - \infty }^{ \infty } dt
\left[  H_{ R }\left( t| \frac{ \hbar }{ i }
\frac{ \delta }{ \delta J( \cdot ) }\right)  -
H_{\tilde{ R }}\left( t | -  \frac{ \hbar }{ i }
\frac{ \delta }{ \delta \tilde{ J }( \cdot )}\right)\right]\right\}\times
\\
\nonumber &&\!\!\!\times\exp\left\{  - \frac{ i }{ \hbar }  \frac{ 1
}{ 2 } \int_{ - \infty }^{ \infty } dt  \int_{  - \infty }^{ \infty
}   dt' \left[  J_{ A }( t )  \mathcal{ D }^{ AB }( t - t' )  J_{ B
}( t' )  + \tilde{ J }_{ A }( t ) \tilde{ \mathcal{ D } }^{ AB}( t -
t')  \tilde{ J }_{ B }( t' )  +\right.\right.
\\
\label{H1.59} && \left.\left. \!\!+ 2 J_{ A }( t )  \bar{ \mathcal{
D } }^{ AB }( t - t' ) \tilde{ J }_{ B }( t' )  \right] (-1)^{
\varepsilon_{ B } }  \right\} N \left(  \exp\left\{  \frac{ i }{
\hbar }  \int_{ - \infty }^{ \infty }  dt \left(  J_{ A }( t )  -
\tilde{ J }_{ A } ( t )  \right)  Z'^{ A }_{in}  \right\}  \right)\!,  
\eeq
where the delocalization $H_{ R }( t | Z( \cdot) )$ was defined
in (\ref{H1.30}), (\ref{H1.31}),
while its anti-chronologic counterpart reads
\beq
\label{H1.60}
H_{ \tilde{ R } }( t | Z( \cdot ) )  =  H( Z_{ \tilde{ R }}( t ) ),         
\eeq
\beq
\label{H1.61}
Z^{A}_{ \tilde{ R } }( t )  =
\left( Q^{ i }( t  +  {\rm sign}( a - b ) 0) \;\!;  P_{ i }( t ) \right),    
\eeq
and the kernels $\mathcal{ D }$ are defined by
\beq
\label{H1.62} \mathcal{ D }^{ AB }( t - t' ) = \frac{ 1 }{ i \hbar }
\left(  Z'^{ A }_{in}  Z'^{ B }_{in}\;\!\theta(t-t') +
Z'^{ B }_{in}  Z'^{ A
}_{in}\;\!\theta(t'-t)(-1)^{\varepsilon_A\varepsilon_B}
-  N(  Z'^{ A }_{in}  Z'^{ B}_{in} )  \right),    
\eeq
\beq
\label{H1.63}
\tilde{ \mathcal { D } }^{ AB }( t - t' ) =
\frac{ 1 }{ i \hbar } \left(   Z'^{ A }_{in}  Z'^{ B }_{in}\;\!\theta(t'-t) +
Z'^{ B }_{in}  Z'^{ A
}_{in}\;\!\theta(t-t')(-1)^{\varepsilon_A\varepsilon_B}
- N(  Z'^{ A }_{in}  Z'^{ B }_{in}  )  \right),   
\eeq
\beq
\label{H1.64}
\bar{ \mathcal{ D } }^{ AB }( t - t' )  =
\frac{1}{ i \hbar }  \left(  N(  Z'^{ A }_{in}  Z'^{ B }_{in})  -
Z'^{ A }_{in}  Z'^{ B }_{in}  \right).    
\eeq

By applying the respective Fourier representations in
$J$ and $\tilde J$ as to the (\ref{H1.59}), one can
obtain the corresponding path-integral representation
for the $(a, b)$-type $in'$ - symbol of the operator
$U( \infty )$,
\beq
\label{H1.65}
U_{(a,b)}( \infty )  =   \int  [ DV ]  [ D\tilde{ V } ]
\exp\left\{  \frac{ i }{ \hbar }  W  \right\}\;\star\:
\exp\left\{  - \frac{ i }{ \hbar }  \tilde{ W }  \right\}.   
\eeq
Here in (\ref{H1.65}):   the $\star$ is the $N_{ (a,b) }$ - symbol multiplication;
the $W$ is given by (\ref{H1.34}); the
$\tilde{ W }$  is given by (\ref{H1.34}) with the replacements:
\beq
\label{H1.66}
Z^{ A }( t )\;\rightarrow \;
\tilde{ Z }^{ A }( t )  =:  Z'^{ A }_{in}  +  \int_{ - \infty }^{  \infty }  dt'
\tilde{ V }^{ C }( t' ) \;\! \omega_{ CB }  \tilde{ \mathcal{ D } }^{ BA }( t'- t ),     
\eeq
\beq
\label{H1.67}
&&\quad\tilde{ \mathcal{ D } }^{ AB}( t - t' )  =  \mathcal{ D }^{ AB }( t' - t ), \\   
\label{H1.68}
&&Z^{ A }_{ R } \;\rightarrow \;  \tilde{ Z }^{ A }_{ \tilde{ R } },  \quad  
 \; J_{ A }( t ) \; \rightarrow \;   \tilde{ J }_{ A }( t ).    
\eeq
The relation
\beq
\label{H1.70}
\qquad\qquad \star =
\exp\left\{  -  \frac{ \overleftarrow{\partial } }{ \partial Z'^{ A }_{in} }\;\!
i \hbar \;\! \bar{ \mathcal{ D } }^{ AB }
\frac{ \overrightarrow{ \partial } }{ \partial Z'^{ B }_{in} }  \right\}
\eeq
has been used when deriving (\ref{H1.65}), with $\bar{ \mathcal{ D } }^{ AB }$
 given by (\ref{H1.64}) where $t$ and $t'$ dependence omitted \cite{BFrad}.
\\

\section{Conclusion}

In the present paper,  we have considered the general aspects of
quantum-mechanical description of dynamical evolution. It is the
general feature that the typical path-integral solutions of the
quantum dynamics depend actually on the particular choice of the
normal form (symbol) of the operators used. The latter dependence
has the two aspects: the specific form of the boundary conditions as
for the virtual trajectory, and the specific form of the boundary
terms in the action. In the present paper, we have determined both
the mentioned aspects, as demonstrated explicitly in the formulae
(\ref{H1.34}) - (\ref{H1.36}) and (\ref{H1.55}) - (\ref{H1.56}).
\\

\section*{Acknowledgments}
\noindent The authors  would like  to thank Klaus Bering of Masaryk
University for interesting discussions. The work of I. A. Batalin is
supported in part by the RFBR grant 17-02-00317. The work of P. M.
Lavrov is supported by the Ministry of Education and Science of
Russian Federation, grant  3.1386.2017 and by the RFBR grant
16-52-12012.
\\

\appendix
\section*{Appendix A. Antisources in gauge theories}
\setcounter{section}{1}
\renewcommand{\theequation}{\thesection.\arabic{equation}}
\setcounter{equation}{0}

Here in this Appendix A, we extend the above consideration
naturally, as to the general case of gauge theories, whose original
Hamiltonian $H$ does commute with a Fermion nilpotent BFV-BRST charge
operator $Q$,
\beq
\label{HA.1}
[  H,  Q  ]  =  0,  \quad   Q^{2}  =  \frac{ 1 }{ 2 }  [  Q,  Q  ]  = 0.  
\eeq
In this case, we modify the equation (\ref{H1.2}) as
\beq
\label{HA.2}
i \hbar \frac{ \partial U }{ \partial t }  =  \left(  -  J_{ A } Z^{  A }
-J^*_{ A}  ( i \hbar )^{-1}  [ Z^{ A }, Q ]  \right)  U,    
\eeq
by introducing  new antisources,\!\!
\footnote{In fact, these objects are the same as
the antifields introduced in \cite{BV,BV1}.}
\beq
\label{HA.3}
J^*_{ A }( t ), \quad
\varepsilon( J^*_{ A } )  =  \varepsilon_{ A }  +1.    
\eeq
Then, the equation (\ref{H1.4}) becomes
\beq
\label{HA.4}
i \hbar \frac{ \partial Z'^{ A } }{ \partial t }  =  [  Z'^{ A },
H'  -  J_{ B } Z'^{ B }  -
J^*_{ B }  ( i \hbar )^{-1}  [  Z'^{ B },  Q' ]  ],     
\eeq so that the equation of motion holds as for the $Q'$,
\beq
\label{HA.5} i \hbar \frac{ \partial Q' }{ \partial t }  =
[J_{ A }  Z'^{ A },  Q'  ],     
\eeq
where we have used (\ref{HA.1}). The equation (\ref{H1.9}) remains valid with the
new $U$ and $Z'^{ A }$. There is also the new equation as to hold,
\beq
\label{HA.6}
\frac{ \hbar }{ i }  U^{-1}( t )  \frac{ \delta }{ \delta J^*_{ A }(
t' ) }  U( t )   =
\theta( t - t' ) ( i \hbar )^{-1}  [ Z'^{ A },  Q' ] ( t' ).   
\eeq
It follows then from the (\ref{HA.5}), (\ref{HA.6}) that the dynamical change of
the BFV-BRST charge $Q'$ caused with the sources and antisources equals
to
\beq
\label{HA.7}
Q'( t )  -  Q'( - \infty )  =U^{-1}(t)[Q,U(t)]=
- i\hbar\;\! U^{-1}( \infty )  \int_{ - \infty
}^{ t }  dt'
J_{ A }( t' )  \frac{ \delta }{ \delta J^*_{ A }( t' ) }\;\!  U( \infty ).     
\eeq which is an operator valued "ancestor" as to the well-known
quantum master equation. Indeed, by making use of the $in'$-normal
Fourier representation as for the generating operator $U(\infty)$,\footnote{By the way,
the $\tilde{U}(\infty)$ entering (\ref{HA.7a}) is expressed in terms of the
respective spectral density $\bar{U}(\infty)$ as
$$\tilde{U}(\infty)=N\left(\exp\left\{Z'^A_{in}
\left(\frac{\delta}{\delta V^A(\infty)}-\frac{\delta}{\delta V^A(-\infty)}
\right)\right\}\bar{U}(\infty)\right)\!.$$}
\beq
\label{HA.7a}
U(\infty)  =:  \int [ DV ]  \exp\left\{  \frac{ i }{ \hbar }  \int
_{  - \infty }^{ \infty }  dt  J_{ A }( t )
\left( Z^{ A }( t )-Z'^A_{in}\right) \right\}
\tilde{ U }(\infty)  , 
\eeq
where $Z^{ A }( t )$ is defined by (\ref{H1.36}),  we rewrite the (\ref{HA.7}) ( at
$t  = \infty$ )  in the form directly related as to the "naive"
master equation (see also the footnote as for the formula (\ref{HA.7a})),
\beq
\label{HA.7b}
[\tilde{U}(\infty),Q]=\left(\frac{\hbar}{i}\right)^2
 \Delta \tilde { U }(\infty),       
\eeq
with  the $\Delta$
being a functional "odd Laplacian",
\beq
\label{HA.7c}
\Delta  =: - \int_{ - \infty }^{ \infty }  dt \;\! ( -1)^{ \varepsilon_{
A } }
\left( \frac{ \partial }{ \partial t }  \frac{ \delta }{ \delta V^{ A }(
t ) }  \right)  \frac{ \delta }{ \delta J^*_{ A }( t ) }.   
\eeq Let $|\Phi \rangle , |\Phi'\rangle$  be two physical states
annihilated by the Hermitian operator $Q$, \beq \label{HA.7d} Q
|\Phi \rangle  = 0, \quad  Q |\Phi' \rangle  =  0,
\quad   Q  =  Q^{ \dagger }. 
\eeq
It follows then from (\ref{HA.7b}) that the physical matrix element of the
operator $\tilde{ U}( \infty )$ between the two states is annihilated by the
$\Delta$,  (\ref{HA.7c}),
\beq
\label{HA.7e}
\Delta   \langle \Phi'| \tilde{ U }( \infty ) | \Phi \rangle  =  0.  
\eeq

Now, consider in short an $Sp(2)$ extension of the main construction
above. In the latter case, we have an $Sp(2)$
 vector
valued Fermion BFV-BRST charge operator \cite{BLTh1,BLTh2,BLTh3},
\beq
\label{HA.8}
[  H,  Q^{ a }  ]  =  0,  \quad  Q^{ a} Q^{ b }  +  (  a \leftrightarrow b  )  =  [
Q^{ a },  Q^{ b }  ]  =  0.   
\eeq
The equation (\ref{HA.2}) now
modifies  as to become
\beq
\label{HA.9}
i \hbar \frac{ \partial U }{ \partial t }  =
\big(  -  J_{ A } Z^{ A } -  J^*_{ Aa }  ( i \hbar )^{ -1}  [  Z^{ A
},  Q^{ a } ]  -  J^{**}_{ A }  ( i \hbar )^{ -2}  [  [  Z^{ A },
Q^{ a }  ],  Q^{ b }  ] \frac{ 1 }{ 2 } \varepsilon_{ ba }  \big) \;\! U, 
\eeq
where we have introduced the antisources, $J^*_{Aa}( t ), J^{**}_{ A }(t)$,
\beq
\label{HA.10}
\varepsilon( J^*_{ Aa } )  =  \varepsilon_{ A }  + 1, \quad  
\varepsilon( J^{**}_{ A } )  =  \varepsilon_{ A }.    
\eeq
The equation (\ref{HA.4}) now becomes
\beq
\label{HA.12}
i \hbar \frac{ \partial Z'^{ A } }{
\partial t }  =  [  Z'^{ A },  H'  -  J_{ B }  Z'^{ B } -  J^*_{
Bb }  ( i \hbar )^{-1}  [  Z'^{ B },  Q'^{ b }  ] -  J^{**}_{ B } (
i \hbar )^{-2}  [  [  Z'^{ B },  Q'^{ a }  ],  Q'^{ b }  ]  \frac{ 1
}{ 2 }  \varepsilon_{ ba }  ].   
\eeq
In turn the equation (\ref{HA.7}) rewrites in the form
\beq
\nonumber
&&\qquad\qquad\qquad Q'^{\;\!a}( t )  -  Q'^{\;\!a}( - \infty )=
U^{-1}(t)[Q^a,U(t)]=\\
\label{HA.14}
&&
=\!  - i\hbar\;\! U^{-1}( \infty )
\int_{ - \infty}^{ t }  dt'\!\!\left(\!
J_{ A }( t' )  \frac{ \delta }{ \delta J^*_{ A a}( t' ) }-
\varepsilon^{ab} J^*_{Ab}(t')\frac{\delta}{\delta J^{**}_A(t')}\right) U( \infty ),\!\!
\eeq
where the normalization $\varepsilon^{12}=-\varepsilon_{12}=1$ is used.
The respective
counterpart to the formula (\ref{HA.7b}) reads
\beq
\label{HA.15}
[\tilde{U}(\infty),Q^a]=\left(\frac{\hbar}{i}\right)^2
\Delta^a_{+} \tilde { U }(\infty),       
\eeq
where
\beq
\label{HA.16}
&&\qquad\qquad\qquad\qquad\qquad\qquad\qquad \Delta^a_{+}=:\Delta^a+\frac{i}{\hbar}V^a, \\
\label{HA.17}
&&\!\!\!\Delta^a  =: - \int_{ - \infty }^{ \infty }\!\!  dt \;\! ( -1)^{ \varepsilon_{
A } }
\left( \frac{ \partial }{ \partial t }  \frac{ \delta }{ \delta V^{ A }(
t ) }  \right)  \frac{ \delta }{ \delta J^*_{ Aa }( t ) },
\quad V^a=:\varepsilon^{ab} \int_{ - \infty }^{ \infty }\!\! dt\;\!
J^*_{Ab}(t)\frac{\delta}{\delta J^{**}_A(t)}\;\!.   
\eeq
Let $|\Phi\rangle , |\Phi'\rangle$ be two physical states annihilated
by the Hermitian operators $Q^{ a }$,
\beq
\label{HA.16a}
Q^{ a } | \Phi \rangle  =  0,  \quad Q^{ a } | \Phi'\rangle   =   0, \quad
Q^{ a }  =  ( Q^{ a } )^{ \dagger }.    
\eeq
It follows then from (\ref{HA.15}) that the physical matrix element
of the operator $\tilde{ U}( \infty)$
between the two states is annihilated  by the $\Delta^{ a }_{ + }$,  (\ref{HA.16}),
\beq
\label{HA.16b}
\Delta^{ a }_{ + }
\langle\Phi' | \tilde{ U}( \infty ) | \Phi \rangle  =  0.   
\eeq

\end{document}